\documentstyle[12pt]{article}
\def\be{\begin{equation}}
\def\ee{\end{equation}}
\def\ba{\begin{eqnarray}}
\def\ea{\end{eqnarray}}

\title{\bf Weak magnetic dipole moments
in two-Higgs-doublet models}
\author{J.\ Bernab\'eu$^a$, D.\ Comelli$^a$, L.\ Lavoura$^b$,
and Jo\~ao P.\ Silva$^a$ \\
\\
\small $^a$ Departament de F\'\i sica Te\`orica and IFIC\\
\small Universitat Val\`{e}ncia -- CSIC \\
\small E-46100 Burjassot (Val\`{e}ncia), Spain \\
\\
\small $^b$ Universidade T\'ecnica de Lisboa and CFIF \\
\small Instituto Superior T\'ecnico, Edif\'\i cio Ci\^encia (f\'\i sica) \\
\small P-1096 Lisboa Codex, Portugal}
\begin{document}
\maketitle
\begin{abstract}
We investigate the effects of the new scalars
in a two-Higgs-doublet model
on the weak magnetic dipole moments of the fermions
at the $Z$ peak.
Proportionality of the Yukawa couplings to the fermion masses,
and to $\tan{\beta}$,
makes such effects more important for the third family,
and potentially relevant.
For the $\tau$ lepton,
the new diagrams are suppressed by
$v_\tau = 2 \sin^2 \theta_W - 1/2$,
or by powers of $m_\tau/M_Z$,
but may still be comparable to the SM electroweak contributions.
In contrast,
we find that the new contributions for the bottom
quark may be much larger than the SM electroweak contributions.
These new effects may even compete with the gluonic contribution,
if the extra scalars are light and $\tan \beta$ is large.
We also comment on the problem of the gauge dependence of the vertex,
arising when the $Z$ is off mass shell.
We compute the contributions from the new scalars
to the magnetic dipole moments for top-quark production
at the NLC,
and for bottom and $ \tau $ production at LEP2.
In the case of the top,
we find that the SM electroweak and gluonic contributions
to the $Z t {\bar t}$ vertex are comparable.
The new contributions may be of the same order of magnitude
as the standard-model ones,
but not much larger.
\end{abstract}

 \newpage

\section{Introduction}

Thus far,
the nature of the symmetry-breaking sector of the standard model (SM)
remains largely untested,
with a fundamental scalar yet to be found.
In fact,
the number of Higgs multiplets
is not predicted and must be determined by experiment,
just like the number of fermion
families.
Moreover,
the success of supersymmetry coupling-constant unification suggests
that one might have several Higgs doublets,
since the minimal supersymmetric standard model (MSSM) requires two doublets.
In this article we consider a two-Higgs-doublet model (THDM),
which has physical scalars $H^\pm$,
$A^0$,
$H^0$ and $h^0$.

While these particles remain undiscovered,
we may look into the
virtual
effects that they induce in
several phenomena.
For instance,
from the neutral-meson mixings
one can set limits on the flavour-changing scalar vertices.
These limits are so stringent that Glashow and Weinberg \cite{Gla77} and,
independently,
Paschos \cite{Pas77},
introduced the concept of Natural Flavour Conservation,
implemented through a discrete symmetry.
We will assume that one doublet couples to the right-handed up-type quarks,
while the other doublet couples
to the right-handed down-type and charged-lepton fields.
This is the so-called model II,
which includes the scalar sector of the MSSM as a particular case.
In model II,
the Yukawa couplings are proportional to the fermion masses,
so that one should look to the third family for noticeable effects.

Another interesting source of information comes from
the radiative corrections to the gauge couplings of the fermions.
On the one hand,
the $Z$ couples to a fermion $f$ through a vector ($v_f$)
and an axial-vector ($a_f$) coupling:
\be
\frac{i e}{2 s_W c_W}\,
{\bar u}(p_-)
\left[ \gamma^\mu \left( v_f(q^2) - a_f(q^2)\, \gamma_5 \right) \right]
v(p_+)\, .
\label{eq:Zcoupling}
\ee
These couplings occur already at tree level in the SM,
with
\ba
v_f &=& T_{3f} - 2\, Q_f\, s_W^2\, ,
\nonumber\\
a_f &=& T_{3f}\, .
\label{eq:vfaf}
\ea
Here,
$s_W$ and $c_W$ are the Weinberg angle's sine and cosine,
and $Q_f$  and $T_{3f}$ are the fermion's charge
and third component of the weak isospin,
respectively.
The momentum of the $Z$ is $q = p_- + p_+$.
Therefore,
precise measurements are required to disentangle the loop effects.
On the other hand,
the anomalous weak magnetic dipole moments (WMDM) $\mu_f$,
defined to be the couplings of the $Z$ to $f$ of the form
\be
\frac{i e}{2 m_f}\,
{\bar u}(p_-)
\left[ \mu_f(q^2)\, i \sigma^{\mu \nu} q_\nu  \right]
v(p_+)\, ,
\label{eq:Zmagneticmoment}
\ee
arise only at loop level,
making them preferred tools in the search for physics beyond the SM.

In this article,
we compute the WMDMs of the $\tau$ lepton and of the bottom quark
in the model described above.
We separate the SM contributions
from the ones involving new scalar particles,
and study the conditions under which the latter are numerically important.
In particular,
we reproduce the SM computation of the $\tau$-lepton's WMDM
in Ref.~\cite{BeG95}.
We give both analytic formulas for the WMDMs,
and also their numerical values.
The analytic formulas can also be applied to compute the
usual magnetic dipole moments of the fermion with the photon\footnote{In
our notation,
the tree-level coupling of the photon to $f$ is $i e Q_f \gamma^{\mu}$,
while the magnetic dipole moment
is defined as in Eq.~(\ref{eq:Zmagneticmoment}).
The usual definition for the photon coupling \cite{bd},
\be
i\, e\, Q_f\, {\bar u}(p_-)
\left[ \gamma^\mu\, F_1(q^2) +
\frac{i}{2 m_f}\, \sigma^{\mu \nu} q_\nu\, F_2(q^2) \right]
v(p_+)\, ,
\label{eq:photoncouplings}
\ee
yields the following relation with our photon magnetic dipole moment:
$ \mu_f = Q_f F_2 $.}.

The article is organized as follows.
In section 2,
we discuss the WMDM of the $\tau$ lepton in the THDM.
In section 3 we present the results for the bottom quark.
Section 4 is devoted to a discussion of the problem
of the gauge non-invariance of the WMDM
when the $Z$ is off mass shell.
This is crucial for the top quark,
but also for the other fermions,
when they are produced in colliders such as LEP2 or the NLC.
We give some numerical values relevant for these colliders.
We draw our conclusions in section 5.
An appendix contains the analytic expressions
for the $Z$ and $\gamma$ magnetic dipole moments (MDMs),
induced at one-loop level in model II.

\section{The WMDM of the $\tau$}

One expects the $\tau$ lepton to be a leading candidate
in the search for new physics through WMDMs.
This is due to the absence of gluon contributions,
and to the fact that the energy and angular distribution
of the decay products of the $\tau$
can be used to extract information
on the spin density matrix of the $\tau$ pairs.
Several groups have used this method to isolate the dispersive and
absorptive parts of the weak electric dipole moment (WEDM)
\cite{Ber89,BeG94}
and of the WMDM \cite{BeG95,Sti93a,BeG94}.
An analysis,
including all the form factors
in terms of the spin density matrix of the $\tau$,
is given in Ref.~\cite{Sti93b}.
In particular,
the transverse polarization of the $\tau$ (within the collision plane)
measures the real part of the WMDM and the imaginary part of the WEDM,
while the normal polarization (perpendicular to the collision plane)
measures the imaginary part of the WMDM and the real part of the WEDM.
These are gauge-invariant,
observable quantities,
as long as the external particles are on mass shell:
$p_-^2 = p_+^2 = m_\tau^2$,
and  $ q^2 = M_Z^2$.

In the SM,
as well as in model II,
the WEDM is multiloop suppressed
(since it must involve the quark mixing matrix),
and therefore these polarizations measure directly the WMDM.
The calculation of the WMDM of the $\tau$ in the SM
has been performed in Ref.~\cite{BeG95},
who found
\be
\mu^{SM}_\tau(M_Z^2) = (-2.10 - i\, 0.61) \times 10^{-6}\, ,
\label{eq:bernabeu}
\ee
which is well below the expected experimental limit of $10^{-4}$.

A typical diagram contributing to the magnetic
dipole moments is shown in Fig.~1.
We will denote the different diagrams by the particles running
in the loop,
starting with the particle $A$ in between the external fermions,
and proceeding counterclockwise to particles $B$ and $C$.
We denote by $\phi$ the SM Higgs scalar,
and by $\sigma$ ($\chi$) the charged (neutral) Goldstone bosons.

The various contributions are suppressed by different powers of $m_\tau$.
One power is always there due to the definition of $\mu_{\tau}$
[See Eq.~(\ref{eq:Zmagneticmoment})].
Extra powers come from the Yukawa couplings
(in the case of the scalars and of the Goldstone bosons) and,
in the diagrams in which the $Z$ couples in the loop
to two particles with the same spin,
from the mass insertion required to reproduce
the chirality structure of the WMDM.
The latter suppression is avoided when the $B$ and $C$ particles
in Fig.~1 are a scalar and a vector (or vice-versa),
that is,
in the $ \tau \phi Z $ and $ \nu \sigma W $ diagrams.
However,
in the SM as in model II,
the suppression thus avoided
is offset by the proportionality
of the Yukawa couplings to the fermion masses.
That is not the case in the most general THDM,
without Natural Flavour Conservation.

We have performed a complete computation of $\mu_{\tau}$ in the SM,
carefully checking its gauge invariance.
Each of the four sets of diagrams in Fig.~2 is separately gauge invariant
when the external particles are on mass shell.
(One must also use the tree-level relation $ c_W = M_W / M_Z $
to explicitly verify gauge invariance.)
This is also the case for the diagrams involving
the SM Higgs particle $\phi$
(the ones in Figs.~3a and 3b,
with $ h^0 \rightarrow \phi $).
The expressions for the MDM in the 't Hooft--Feynman gauge
are listed in the appendix.
We agree with the results in Ref.~\cite{BeG95},
except for a few signs,
and for the small
$\tau \phi Z$ and $\tau Z \phi$ contributions,
which  do not affect significantly the final numerical results.

For the numerical computations,
we used the following values.
At $ q^2 = M_Z^2 $,
$ \alpha_{em} = 1/128 $ and $ s_W^2 = 0.232 $.
(Notice that,
due to the large cancellation in $ v_{\tau} = 2 s_W^2 - 1/2 $,
some of the numerical results are very sensitive
to the input value of $s_W^2$.)
We use $ M_Z = 91 $ GeV,
$ M_W = c_W M_Z $,
and $ m_{\tau} = 1.777 $ GeV.
We take the neutrino to be massless.
Using this input,
we get the values for the contributions to the $\tau$ WMDM
displayed in Table~1.
\begin{table}[htb]
\centering
\begin{tabular}{|c|c|}
\hline
\hline
$ABC$ &
$ \mu_{\tau}^{ABC} $ \\
\hline
$ \gamma \tau \tau $ &
$ (3.1853 - i\, 1.2713) \times 10^{-7}  $ \\
\hline
$ Z \tau \tau + \chi \tau \tau $ &
$ (4.1316 + i\, 1.9133) \times 10^{-8} $ \\
\hline
$ W \nu \nu + \sigma \nu \nu $ &
$ (- 9.8734 - i\, 5.0748) \times 10^{-7} $ \\
\hline
$ \nu W W + \nu \sigma W + \nu W \sigma + \nu \sigma \sigma $ &
$ -1.4733 \times 10^{-6} $ \\
\hline
Total &
$ - 2.1008 \times 10^{-6} - i\, 6.1548 \times 10^{-7} $ \\
\hline
\hline
\multicolumn{2}{l}{Table 1: Standard-model contributions
to the WMDM of the $\tau$}
\end{tabular}
\end{table}
We have grouped the diagrams into the gauge-invariant classes
shown in Fig.~2.
Our final result agrees with that of Ref.~\cite{BeG95}.

We stress that all the results presented in this article are
based on exact formulas, which we have evaluated using two
completely different programs, as a cross-check.
In many cases, the numerical results may be guessed at from
the analytical expressions in the appendix by taking the relevant
integrals to be dominated by the largest mass scale in the
diagram.
In the appendix, we show explicitly that this `guesstimate' of
the integrals is correct for the $\gamma \tau \tau$ diagram.
Experience shows that it also works for many other diagrams,
making it a valuable tool in understanding the relative magnitudes
of the different contributions.

We now consider the scalar-particle contributions,
both in the SM and in the THDM.
The relevant diagrams are the ones shown in Fig.~3,
and their contributions are given in the appendix.
There are four types of scalar contributions:
those with a neutral scalar and a $Z$ (Fig.~3a),
those with a neutral scalar (Fig.~3b),
those with a neutral pseudo-scalar (Fig.~3c),
and those with a charged scalar (Figs.~3d and 3e).
In the SM, only the first two types of diagrams are present.
They are functions of the Higgs mass which
we display in Figs.~4 and 5, for the
$\tau \phi Z + \tau Z \phi$, and $\phi \tau \tau$ diagrams,
respectively. We have taken all the scalar masses to be greater
than 50 GeV.
One can see that the SM Higgs contributions
are never larger than about 2\% of the non-Higgs contributions.

In model II,
the contribution of diagram 3a to the WMDM
is proportional to
$\cos^2 \alpha + \tan \beta \sin \alpha \cos \alpha$,
for $H^0$,
and to $\sin^2 \alpha - \tan \beta \sin \alpha \cos \alpha$,
for $h^0$,
where $\alpha$ is the mixing angle in the CP-even mass matrix,
and $\tan \beta$ is the ratio between the two vev's.
In the following we shall take
$1 < \tan \beta <70$ and let $\alpha$ take any value.
The proportionality factor is the function plotted in Fig.~4.
We find that this contribution may increase or decrease
the real part of the WMDM by about 80\%,
for a scalar of mass 50 GeV.
Of course,
this diagram does not
originate an imaginary part since,
at $q^2 =M_Z^2$,
no cut can be made.

Similarly,
the amplitude generated by
diagram 3b is proportional to the function of the mass of the scalar
shown in Fig.~5.
The proportionality factors are $(1 + \tan^2 \beta)$
times $\cos^2 \alpha$ for $H^0$,
or times $\sin^2 \alpha$ for $h^0$.
The appearance of $\tan^2 \beta$ makes this contribution
potentially large.
It turns out that the contribution to the real part of the WMDM
is negligible,
but that the contribution to its imaginary part
can make it to about one third of the SM value,
for a scalar of mass 50 GeV.
This happens despite the fact that this diagram has
an extra $m_\tau^2/M_Z^2$ suppression factor.

The contribution from the pseudo-scalar in diagram 3c
is equal to $ \tan^2 \beta $ times
a function of the mass of $A^0$ displayed in Fig.~6.
We see that,
if that mass is 50 GeV and $ \tan^2 \beta $ is large,
this may decrease the imaginary part of the WMDM
by about 25\%.
The influence on the real part of the WMDM is negligible.

Finally we turn to the contribution from the charged scalar
involved in diagrams 3d and 3e.
We show in the appendix that such contributions are,
in general,
linear combinations of $1$,
$\tan^2 \beta$,
and $\cot^2 \beta$.
However,
due to the zero mass of the neutrino,
in this case only the term proportional to $\tan^2 \beta$
is non-vanishing.
The factor of proportionality is displayed in Fig.~7 as
a function of the mass of the charged scalar.
We see that both the real and the imaginary parts of the
charged-scalar contribution
may make it to about $ 1 \times 10^{-6} $,
if the mass of that particle is 50 GeV,
and $ \tan \beta $ is maximal.
This is a contribution of the same order of magnitude as the SM one,
but of the opposite sign.

In conclusion,
we find that all the new diagrams arising in model II
may have an impact on the WMDM of the $\tau$ lepton,
if $\tan \beta \sim 70$ and the mass of the scalar particles is
$\sim 50$ GeV.
They may be comparable to the SM contributions,
but never much larger than them.
This is partly due to the masslessness of the neutrino,
which eliminates some terms from the general expressions,
and partly due to the proportionality to the vector
coupling of the tau,
$ v_\tau $,
which is approximately $ -0.036 $.
This will not be the case for the quarks.

\section{The WMDM of the bottom quark}

The measurement of the WMDM of the bottom quark is more
problematic since its polarization is affected by hadronization.
Still, one may look for its influence on the
$Z \rightarrow b \bar{b}$ observables.
Currently, this yields bounds of the order of $10^{-2}$
\cite{Zbbbar}.

Another possibility is brought about by the fact that
the initial spin of the bottom quark is retained in the
polarization of the $\Lambda_b$ baryons that are produced directly
\cite{Clo92,Man92},
although there might be substantial depolarization induced
by the $\Lambda_b$'s produced from the decay of heavier
b-baryons \cite{Clo92,Fal94}.
This polarization can be studied
by measuring the energy spectra of the charged-leptons
\cite{cleptonspectra},
or the neutrinos
\cite{neutrinospectra},
produced in the semileptonic decays of the $\Lambda_b$.

In the calculation of the WMDM of the bottom quark,
we have used the values of $M_Z$,
$M_W$,
$s_W$ and $\alpha_{em}$ given in the previous section.
In addition,
we take $ m_b = 5 $ GeV,
$ m_t = 174 $ GeV,
and $\alpha_S (M_Z) = 0.117$.
The values obtained are listed in Table~2.
\begin{table}[htb]
\centering
\begin{tabular}{|c|c|}
\hline
\hline
$ABC$ &
$ \mu_b^{ABC} $ \\
\hline
$ g b b $ &
$ (3.5764 - i\, 1.9382) \times 10^{-4}  $ \\
\hline
$ \gamma b b $ &
$ (1.9900 - i\, 1.0785) \times 10^{-6}  $ \\
\hline
$ Z b b + \chi b b $ &
$ (3.0041 + i\, 1.3393) \times 10^{-6} $ \\
\hline
$ W t t + \sigma t t $ &
$ - 2.2064 \times 10^{-6} $ \\
\hline
$ t W W + t \sigma W + t W \sigma + t \sigma \sigma $ &
$ - 6.4791 \times 10^{-6} $ \\
\hline
Total &
$ (3.5394 - i\, 1.9356) \times 10^{-4} $ \\
\hline
\hline
\multicolumn{2}{l}{Table 2: Standard-model contributions
to the WMDM of the bottom}
\end{tabular}
\end{table}
Contrary to what happened in the case of the $\tau$,
for the case of the bottom,
all the gauge-invariant electroweak contributions
are of the same order of magnitude,
except for the Higgs contributions,
which amount to less than 1\% of the overall electroweak result.
However,
the WMDM of the bottom quark is dominated by the gluon diagram,
shown in Fig.~8,
which is about two orders of magnitude larger than the
electroweak ones.

Still,
the new scalars in model II may be important since the
two factors limiting their impact on the WMDM of the $\tau$
lepton are not present in the case of the bottom quark.
In fact,
the Yukawa-coupling factors in the vertices
are now often factors of enhancement instead of suppression,
because of the large coupling of the top,
and also of the bottom
in the large-$\tan{\beta}$ limit.
In addition,
the vector coupling
$ v_b = 2/3 \sin^2 \theta_W - 1/2 \approx - 0.345 $
is an order of magnitude larger than $v_\tau$.

The shape of the contributions from diagrams 3a and 3b
as functions of the scalar masses are similar to those of
Figs.~4 and 5, although the numerical values are
different.
We find that the diagrams in Fig.~3a
may increase or decrease the real part of the WMDM by some 35\%,
if one of the scalars of the THDM is very light
while the other one is very heavy,
and if $\tan \beta$ is very large.
Similarly,
the $ h^0 b b $ or $ H^0 b b $ diagrams in Fig.~3b
may be important for the imaginary part of the WMDM,
increasing its SM value by about 70\%
if $\tan \beta$ is large and the scalars are light.

As before,
the pseudo-scalar contribution is proportional to
$\tan^2 \beta$ and,
if $\tan^2 \beta = 5000$ and $m_{A^0} = 50$ GeV,
that contribution is maximal,
reaching $ (-3 + 9 i) \times 10^{-5} $.

Finally,
we look at the contributions from the charged scalar $H^{\pm}$.
These are real since we are assuming that the mass of
the charged scalar is larger than $M_Z/2$.
As we have mentioned above,
these contributions are linear combinations of $1$,
$\tan^2 \beta$,
and $\cot^2 \beta$.
We find that these contributions are,
at best,
two orders of magnitude smaller than the
QCD contribution,
for $\tan^2 \beta = 5000$
or $\tan^2 \beta = 1$
(in which case it is dominated by the
term proportional to 1).

So,
contrary to what happened for the $\tau$ lepton,
for the bottom quark only the neutral-scalar
and pseudoscalar contributions
may change the SM value for the WMDM appreciably.
One should note the remarkable fact that these
new scalar contributions may compete with the strong
gluonic corrections.

\section{Magnetic dipole moments at $q^2 > M_Z^2$}

The CDF and D0 Collaborations \cite{CDF}
have shown that the top quark is quite heavy.
Because of this fact,
problems arise in the definition of an appropriate WMDM for the top,
since in order to produce a top pair the $Z$ must be off mass shell.
The root of the problem lies in the gauge dependence of
any form factor arising in the Lorentz decomposition of the
$Z \bar{f} f$ and $\gamma \bar{f} f$ vertices,
when the gauge boson ($Z$ or $\gamma$) is off mass shell \cite{Lee}.
Indeed,
the MDMs are unequivocably defined
(gauge-invariant and observable)
only when the incoming gauge boson is on mass shell
($q^2=0$ for the $\gamma$,
and $q^2= M_Z^2$ for the $Z$),
{\it i.e.},
when the physical process is dominated by the
single-gauge-boson exchange.

Of course,
this problem does not exist solely with the top quark.
In particular,
it occurs whenever one studies
the radiative corrections to fermion-pair production
at colliders with energy beyond the $Z$ mass,
such as LEP2 or the NLC.

Recently,
the pinch technique \cite{Cor77}
has been used to construct gauge-independent MDMs \cite{Pap94}.
This technique consists in reorganizing
the usual Feynman-diagram expressions
into portions that are manifestly gauge independent.
One extracts from the box diagrams those gauge-dependent
pieces which are kinematically equivalent to the
$\gamma \bar{f} f$ or to the $Z \bar{f} f$ vertex corrections.
Those pieces offset the gauge-noninvariance of the vertex corrections.

One may now calculate these quantities in any gauge
as long as the pinch contributions are consistently identified
in the box and vertex graphs.
However,
in this particular case,
the computation in the 't Hooft--Feynman gauge is most convenient
since, in that gauge,
the MDMs do not receive any contributions from the box diagrams.
Indeed,
such contributions from the box diagram
arise from the longitudinal terms
in the gauge-boson propagators,
which are absent in this gauge.
This means that,
in the 't Hooft--Feynman gauge,
the MDMs only receive contributions from the vertex corrections themselves.
Thus,
the fact that kinematically the top quark production requires
$q^2 > 4 m_t^2$,
would be circumvented by a judicious definition of the form factors
when the vector boson is off mass shell.

However,
other authors \cite{Den95}
have claimed that the improved off-shell vertex functions
are not uniquely defined.
The basic problem is that one can still shift gauge-independent
pieces between the various diagrams.
This ambiguity means that the gauge-invariant quantities
obtained from the pinch technique are not observable
by themselves.
Still, they may be useful in determining which new physics
effects may be important, and where to look for them.

Conscious of these problems,
we have computed the contributions to the
MDMs of the top quark with the $\gamma$ and with the $Z$
at $\sqrt{q^2} = 500$ GeV
(the center-of-mass energy expected
for the next $e^+\ e^-$ linear collider,
NLC)
following the prescription of Ref.~\cite{Pap94}.
We have also computed the MDMs of the bottom
and of the the $\tau$ at $\sqrt{q^2} = 200$ GeV
(relevant for LEP2).
We do this in order to have an idea of the
sensitivity of those MDMs to the new physics in the THDM.
Our calculations are to be considered
as preliminaries to a more general work
in which the box diagrams should also be taken into account,
in the computation of suitably defined physical observables.

\subsection{The top quark at the NLC}

For the NLC,
we have taken $\alpha_S = 0.096$,
$\alpha_{em} = 1/127$ and $\sin^2 \theta_W = 0.240$.
These values are obtained
by a SM renormalization-group running of these parameters
from their measured values at $q^2 = M_Z^2$.
In Table~3,
we present the SM contributions to the WMDM in the
$ Z \bar{t} t $ vertex at $\sqrt{q^2} = 500$ GeV.
\begin{table}[htb]
\centering
\begin{tabular}{|c|c|}
\hline
\hline
$ABC$ &
$ \mu_t^{ABC} $ \\
\hline
$ g t t $ &
$ (-2.6170 + i\, 4.5493) \times 10^{-3}  $ \\
\hline
$ \gamma t t $ &
$ -7.1549 \times 10^{-5} + i\, 1.2438 \times 10^{-4}  $ \\
\hline
$ Z t t + \chi t t $ &
$ -2.4585 \times 10^{-4} - i\, 1.4502 \times 10^{-3} $ \\
\hline
$ W b b + \sigma b b $ &
$ -4.2241 \times 10^{-4} + i\, 1.2600 \times 10^{-3} $ \\
\hline
$ b W W + b \sigma W + b W \sigma + b \sigma \sigma $ &
$ -2.2436 \times 10^{-3} + i\, 8.7342 \times 10^{-4} $ \\
\hline
Total &
$ (-5.6004 + i\, 5.3569) \times 10^{-3} $ \\
\hline
\hline
\multicolumn{2}{c}{Table 3:
Standard-model contributions to the WMDM of the top,} \\
\multicolumn{2}{c}{at $\sqrt{q^2} = 500$ GeV}
\end{tabular}
\end{table}
Notice that,
contrary to the naive expectation,
the electroweak sector gives contributions comparable to the
gluonic one.
Qualitatively, this difference with respect to the case of
the bottom quark is due to the fact that here the largest mass scale
that may dominate the integrals is never much larger than $m_t^2$,
while the mass of the bottom satisfies
$m_b^2 \ll M_Z^2, m_t^2$.

Now we can discuss the effect induced by the new scalars
of the THDM.
If the new scalar particles can be produced directly at
$\sqrt{q^2}=500$ GeV,
their effect on the MDMs will be to generate new
non-vanishing imaginary parts due to unitarity.
We are more interested in studying a possible scenario,
in which they only have virtual effects.
Therefore,
we are looking for the possibility of sizeable
effects induced by a charged scalar with mass larger
than 250 GeV (meaning that no $H^+ H^-$ event has been detected)
and neutral scalars bounded by the following constraints:
$m_{h^0, H^0} \geq 409$ GeV (no scalar--$Z$ event);
and $m_{h^0, H^0}+m_{A^0} \geq 500$ GeV
(no scalar--pseudoscalar event).
The formulas for the various contributions are listed in the
appendix.
Notice that for the top quark there is a
$\tan{\beta} \leftrightarrow \cot{\beta}$
interchange with respect to the formulae for a fermion with $T_{3f}=-1/2$.

We find that the contribution from diagram 3a,
which is real since we have taken the scalar masses to be greater
than 409 GeV to avoid direct production,
is now maximal for $\tan{\beta} \sim 1$
but may only reach 10\%
of the SM value for scalar masses close to 409 GeV.
This value decreases very rapidly as one goes away from
409 GeV into higher masses,
as a consequence of the threshold effect
that exists close to production.

Similarly,
diagrams 3b and 3c take on their maximum values for
$\tan{\beta} \sim 1$.
This will limit their impact,
although the extra $m_t^2/M_Z^2$ prefactors
enhance these contributions.
We find that these contributions
never go beyond 15\% of the SM result,
in their imaginary and real parts.

Finally,
the charged-scalar contributions permit a test of both
regimes:  $\tan{\beta} \gg 1$ and $\tan{\beta} \sim 1$.
In the first regime the dominant piece is the one proportional
to $\tan{\beta}$,
which may reach two thirds of the SM value in its real part,
but only 12\% of the SM value in its imaginary part.
For the  $\tan{\beta} \sim 1$ regime,
one may reach around 20\% of the SM real part and
decrease the SM imaginary part by about 15\%.
These numbers are obtained for a charged scalar with mass 250 GeV,
but decrease very slowly for higher values of the mass.

The SM contributions to the MDM coupling of the $ \gamma \bar{t} t $
vertex at $\sqrt{q^2} = 500$ GeV
are listed in Table~4.
\begin{table}[htb]
\centering
\begin{tabular}{|c|c|}
\hline
\hline
$ABC$ &
$ \mu_t^{ABC} $ \\
\hline
$ g t t $ &
$ -8.2790 \times 10^{-3} + i\, 1.4392 \times 10^{-2}  $ \\
\hline
$ \gamma t t $ &
$ (-2.2635 + i\, 3.9349) \times 10^{-4}  $ \\
\hline
$ Z t t + \chi t t $ &
$ -6.4295 \times 10^{-5} - i\, 3.5312 \times 10^{-3} $ \\
\hline
$ W b b + \sigma b b $ &
$ (-1.4404 + i\, 4.2711) \times 10^{-4} $ \\
\hline
$ b W W + b \sigma W + b W \sigma + b \sigma \sigma $ &
$ -1.4978 \times 10^{-3} + i\, 7.0128 \times 10^{-4} $ \\
\hline
Total &
$ (-1.0212 + i\, 1.2383) \times 10^{-2} $ \\
\hline
\hline
\multicolumn{2}{c}{Table 4:
Standard-model contributions to the MDM of the top with the $\gamma$,} \\
\multicolumn{2}{c}{at $\sqrt{q^2} = 500$ GeV}
\end{tabular}
\end{table}
The gluonic contribution for this vertex is enhanced
with respect to the previous one by $Q_t/x_t \approx 3.16$.
This fact makes it the dominant contribution to this MDM.

Again the contributions from the new neutral scalars may,
at most,
reach 20\% of the SM value,
being more important for the imaginary part of the MDM.
(Recall that the photon does not have a contribution
from diagram 3a.)
In contrast,
the charged-scalar contributions may reach two thirds
of the  real part of the SM value,
for $\tan^2 \beta = 5000$ and a charged-scalar mass near the
threshold for $H^+ H^-$ production.
These contributions are not as large for $\tan^2 \beta \sim 1$
where one can only reach 15\% of the real part of the SM value.

\subsection{The $\tau$ lepton and the bottom quark at LEP2}

The calculations of the WMDM and the electromagnetic MDM for
the bottom and $\tau$ at LEP2 follow those in the previous
section.
At LEP2,
with $\sqrt{q^2} = 200$ GeV,
we use $\alpha_S = 0.106$,
$\alpha_{em} = 1/127.5$ and $\sin^2 \theta_W = 0.236$.
We remind the reader of the ambiguity in defining appropriate
MDMs off the intermediate vector boson mass shell,
and report our results only qualitatively.
We take these as a hint of where to look for important
virtual effects of the new scalars.
We shall take the charged-scalar masses to be greater than
100 GeV, together with the constraints
$m_{h^0, H^0} \geq 109$ GeV and
$m_{h^0, H^0}+m_{A^0} \geq 200$ GeV.

Phenomenologically,
the $\tau$ may be more interesting,
due to the possibility of a clean measurement of its polarization.
The SM values are:
$ (-1.1142 - i\, 3.4060) \times 10^{-6} $ for the WMDM,
and $ (1.3347 - i\, 2.6968) \times 10^{-6} $ for the MDM with the photon.
These values are comparable with those obtained for the
WMDM at $M_Z^2$.

Due to the arguments explained in section 2,
the contributions from the new scalars in the THDM only affect
the MDMs in the region of large $\tan \beta$.
In the case of the WMDM,
the contribution from diagram 3a may be very large.
For a mass of the scalar near the threshold of 110 GeV,
one gets two times the SM real part.
All the other contributions are negligible,
except for the charged-scalar ones which can reach 25\% (8\%)
of the real (imaginary) part of the SM result,
but with the opposite sign.
This occurs for $\tan^2 \beta = 5000$
and charged scalar masses of 100 GeV.

For the photon vertex all the diagrams may give relevant contributions,
for $\tan^2 \beta = 5000$ and light scalars.
Diagram 3b may have an imaginary piece as large as 40\% of the SM one.
The pseudo-scalar correction may reduce
the real part of the SM result by 60\%
and its imaginary part by 30\%.
Finally, the charged scalar may increase the real part by 25\%.

For the bottom quark at LEP2,
we have found the following  results:
the standard-model WMDM is
$ (7.6143 - i\, 4.0685) \times 10^{-5} $,
while the standard-model MDM with the $\gamma$ is
$ (5.8639 -\,i\, 3.4638) \times 10^{-5} $.
These values are about a factor of five smaller than
the WMDM at $q^2=M_Z^2$.
Both MDMs are dominated by the QCD correction to the vertex.

In the case of the $Z$ vertex,
there is a substantial contribution to the real part
coming from diagram 3a,
which may amount to three times the gluonic one,
for a neutral scalar close to 109 GeV.
On the other hand,
the dominant new scalar contributions to the imaginary
part come from diagram 3b,
which may increase the gluonic value by 70\%,
and diagram 3c, which may decrease it by 50\%.
The pseudo-scalar diagram may decrease the real part of the WMDM by 25\%.
All these results are for $\tan^2 \beta = 5000$.

For the photon vertex, the largest corrections to the
imaginary part that one may obtain come from
diagram 3b (increase of 70\%) and diagram 3c (decrease of 50\%),
for large $\tan \beta$.
Also,
the pseudo-scalar diagram may decrease the real part by 25\%.

\section{Conclusions}

We have performed a complete analysis of the magnetic dipole
moments of the fermions in the SM and compared them to the
corrections that may be induced by the virtual scalars in
a THDM with a discrete symmetry, the so-called model II.
For the $\tau$ lepton, the masslessness of the neutrino
and the smallness of $v_\tau$ suppress the new contributions
which may, nevertheless, be as large as the SM electroweak
corrections.
For the bottom quark, these suppressions disappear and
the new physics yields MDMs much larger than the electroweak
ones.
They may even compete with the gluonic correction.
We have also presented the results for the MDMs
defined by the pinch technique,
which are relevant for the top quark produced at the NLC,
and for the $\tau$ lepton and for the bottom quark produced at LEP2.
This definition is ambiguous,
but it is useful in identifying candidate situations
in which to look for new physics.
In so doing we find some interesting results.
The pure electroweak radiative corrections to the
$Z t {\bar t}$ vertex at NLC are as large as the gluonic
corrections, although that is not the case for the
$\gamma t {\bar t}$ vertex.
The diagrams in Fig.~3a can give contributions to the real part
of the WMDM of the bottom and $ \tau $ two or three times larger
than the SM values.
In most other cases,
precise measurements of the magnetic dipole moments
would be required to disentangle the new effects.
Those effects are maximal for light scalars,
allowing for tests of the region of parameter space with large $\tan \beta$,
both in the case of the $\tau$ and of the bottom.

\subsection{Acknowledgements}

We are indebted to G.\ Rodrigo
for sharing with us his expertise
on Passarino--Veltman functions.
We are also grateful to J.\ Papavassiliou
for clarifying exchanges on the pinch technique,
and for correcting a few mistakes in Ref.\ \cite{Pap94}.
This work has been supported in part by CICYT under grant AEN 93-0234.
The work of D.\ C.\ was funded by the Spanish Comisi\'on
Interministerial de Ciencia y Tecnolog\'\i a.
The work of J.\ P.\ S.\ was funded by the European Union under the program
of Human Capital and Mobility.
J.\ P.\ S.\ is indebted to G.\ C.\ Branco and
the Grupo Te\'orico de Altas Energias for their kind hospitality.



\appendix

\section{Formulas for the magnetic moments}

In this appendix we give the magnetic dipole moments
of a fermion $f$ induced at one-loop level in the THDM.
We present the results in the 't Hooft--Feynman gauge.
They hold for any fermion,
and are written for both the $Z$ and $\gamma$ magnetic moments,
for any value $q^2$ of their squared momentum.
However,
some of the results are gauge-invariant
only if the gauge bosons are on mass shell,
as we have found by computing them in a general 't Hooft gauge.

For simplicity,
we introduce the definition
\be
\mu^{ABC} = \frac{\alpha_{em} m_f^2}{4 \pi}\, b^{ABC}\, ,
\label{eq:bABCdefinition}
\ee
where $A$,
$B$ and $C$ are the particles in the loop,
in the order displayed in Fig.~1.
Moreover,
$x_f$ and $y_f$ are defined as
\be
x_f = \left( \begin{array}{c}
		v_f / (2 s_W c_W) \\
		Q_f
	     \end{array}  \right)\, ,
\hspace{5mm}
y_f = \left( \begin{array}{c}
		a_f / (2 s_W c_W) \\
		0
	     \end{array}  \right)\, ,
\label{eq:xydefinition}
\ee
where the upper lines hold
when the exterior gauge boson is the $Z$,
while the lower lines hold for the $\gamma$
MDMs.
The quantities $ v_f $ and $ a_f $ are given in Eq.~(\ref{eq:vfaf}).
The letter $f$ refers to the fermion whose
MDM is being calculated,
and $i$ to its SU(2)$_L$-doublet partner.

In order not to clutter our formulas,
we will omit possible mixing-matrix elements,
notably in the quark sector.
The inclusion of those mixing-matrix elements is clearly trivial.
Furthermore,
our formulas are valid for any fermion.
In the text,
they have only been used to compute the MDMs
of the fermions of the third family,
because those are the ones for which the contributions
from the extended scalar sector are larger.

We use for the integrals the notation of Ref.~\cite{BeG95}:
\ba
  &  &
\left[ I_{00} ;\, I_\mu ;\, I_{\mu \nu} \right]
(m_A, m_B, m_C, q)
\nonumber\\*[3mm]
  &  &
= \int \frac{d^4 k}{i \pi^2}\,
\frac{ \left[1 ;\, k_\mu ;\, k_\mu k_\nu \right]}{ (k^2-m_A^2)
[(k-p_-)^2-m_B^2] [(k+p_+)^2-m_C^2]}\, .
\label{eq:integrals}
\ea
They are decomposed as
\ba
I^\mu & = &
(p_- - p_+)^\mu\, I_{10} + (p_- + p_+)^\mu\, I_{11}\, ,
\nonumber\\*[3mm]
I^{\mu \nu} & = & (p_+^\mu p_+^\nu + p_-^\mu p_-^\nu)\, I_{21}
+ (p_+^\mu p_-^\nu + p_-^\mu p_+^\nu)\, I_{22}
\nonumber\\*[1mm]
   &   &
+\, (p_+^\mu p_+^\nu - p_-^\mu p_-^\nu)\, I_{2, -1}
+ g^{\mu \nu}\, I_{20}\, .
\label{eq:decomposition}
\ea
These functions are trivially related to the
Passarino--Veltman functions \cite{PaV79}.
If $m_B=m_C$,
$I_{11}$ and $I_{2,-1}$ vanish.
Also,
notice that,
from $ I_{\mu \nu} $,
only the combination $ I_{21} - I_{22} $
appears in the expressions for the MDMs.

We have obtained the numerical values of these integrals
in two independent ways.
On the one hand,
FeynCalc was used to reduce them to functions of the scalar integrals $A_0$,
$B_0$ and $C_0$,
for which there are explicit formulas.
The results were checked against the ones obtained
with the aid of the ff routines of van Oldenborgh \cite{oldenborgh}.

\subsection{SM contributions, without the Higgs scalar}

In this subsection we include the contributions
from diagrams involving only SM particles,
except the Higgs scalar.
The latter contributions will be included in the next section,
where we discuss the scalar sector of the THDM.
We find
\be
b^{\gamma f f} =
8\, x_f\, Q_f^2
\left[ I_{21}-I_{22}-I_{10} \right](0,m_f,m_f,q)\, .
\label{eq:gammaff}
\ee
For quarks only,
there is a similar contribution with a gluon $g$
instead of the photon:
\be
b^{g f f} =
8\, x_f\, C_F\, \frac{\alpha_S}{\alpha_{em}}\,
\left[ I_{21}-I_{22}-I_{10} \right](0,m_f,m_f,q)\, ,
\label{eq:gluonff}
\ee
where $ C_F = 4/3 $.
For these contributions,
the integrals can be computed analytically
and one obtains the explicit formula
\ba
\left[ I_{21} - I_{22} - I_{10} \right](0,m_f,m_f,q)
   &  =  &
\frac{1}{4 q^2}\, \int_0^1  \frac{dy}{y(y-1)+m^2/q^2}
\nonumber\\*[3mm]
   &  =  & \frac{1}{2 q^2 \delta}\,
\left\{ \begin{array}{ll}
		\log{\frac{\delta - 1}{\delta + 1}}\, , &
		\mbox{for}\ q^2 < 0\, ;\\*[2mm]
		\log{\frac{1-\delta}{1+\delta}} + i \pi\, , &
		\mbox{for}\ q^2 > 4 m_f^2\, ,
	\end{array}
\right.
\label{eq:schwinger}
\ea
where
\be
\delta = \sqrt{1 - \frac{4 m_f^2}{q^2}}\, .
\label{eq:deltadefinition}
\ee
The region $0 < q^2 < 4 m_f^2$ is unphysical.
In the limit $ q^2 = 0 $ one obtains $1/(4 m_f^2)$,
reproducing the Schwinger term.
Thus,
this integral is dominated by the largest mass scale:
it is of order $1/m_f^2$ when $4 m_f^2 \gg |q^2|$,
and of order $1/q^2$ when $q^2 \gg 4 m_f^2$.
This is a common feature of these integrals,
which is crucial to obtain the correct decoupling limits.

For the diagrams in Fig.~2b we find
\ba
b^{Z f f} &=&
\frac{2}{s_W^2 c_W^2}
\left[ x_f v_f^2 \left( I_{21}-I_{22}-I_{10} \right)
+ x_f a_f^2 \left( I_{21} - I_{22} - 5 I_{10} + 2 I_{00} \right)
\right.
\nonumber\\
   &   &
   \hspace{12mm}
\left.
+ 2 y_f v_f a_f \left( I_{21} - I_{22} - 3 I_{10} + I_{00} \right)
\right](M_Z,m_f,m_f,q)\, ,
\nonumber\\*[3mm]
b^{\chi f f} &=&
\frac{x_f m_f^2}{s_W^2 M_W^2}
\left[ I_{21}-I_{22} \right](M_Z,m_f,m_f,q)\, .
\label{eq:Zff}
\ea
The sum of these two contributions is gauge-invariant.

Similarly,
the sum of the two contributions
\ba
b^{W i i} &=&
2\, \frac{x_i+y_i}{s_W^2}\,
\left[ I_{21} - I_{22} - 3 I_{10} + I_{00} \right](M_W,m_i,m_i,q)\, ,
\nonumber\\*[3mm]
b^{\sigma i i} &=&
\frac{1}{s_W^2 M_W^2}
\left\{
\left[ m_f^2 (x_i+y_i) + m_i^2 (x_i-y_i) \right]
\left( I_{21}-I_{22}-I_{10} \right)
\right.
\nonumber\\
   &  &
\left.
	\hspace{17mm}
	+\, 2\, x_i\, m_i^2\, I_{10}
\right\}(M_W,m_i,m_i,q)\, ,
\label{eq:Wii}
\ea
is gauge-invariant.

Finally,
\ba
b^{i W W} &=& {\cal S}\,
\left(\begin{array}{c}
	\cot \theta_W \\
		1
	\end{array}\right)
\frac{1}{s_W^2}
\left[ 2 I_{21} - 2 I_{22} + I_{10}
\right](m_i,M_W,M_W,q)\, ,
\nonumber\\*[3mm]
b^{i \sigma \sigma} &=& {\cal S}\,
\left(\begin{array}{c}
	\cot{(2 \theta_W)}\\
		1
	\end{array}\right)
\frac{1}{s_W^2 M_W^2}
\left[  \left(
	   m_f^2 + m_i^2
	\right)
	\left(I_{21}-I_{22}-I_{10}\right)
\right.
\nonumber\\
  &  &
\left.
	\hspace{45mm}
	   + m_i^2 \
	\left(I_{00}-2I_{10}\right)
\right](m_i,M_W,M_W,q)\, ,
\nonumber\\*[3mm]
b^{i \sigma W} &=& b^{i W \sigma} = {\cal S}\,
\left(\begin{array}{c}
	- \tan{(\theta_W)}\\
		1
	\end{array}\right)
\frac{1}{2 s_W^2}
I_{10}(m_i,M_W,M_W,q)\, .
\label{eq:iWW}
\ea
Once again,
only the sum of these four contributions is gauge invariant.
Moreover,
gauge invariance in this case only occurs if,
either $q^2 = M_Z^2$ and one uses the upper line for the coefficients
(for the case of a $Z$ as external gauge boson),
or $q^2 = 0$ and one uses the lower line
(for the case of a $\gamma$ as external gauge boson).
This is the reason behind the
gauge noninvariance of the MDMs
at arbitrary $q^2$.
Gauge invariance of a physical observable
(such as the $e^+ e^- \rightarrow f {\bar f}$ cross section)
is restored if we add to Eqs.~(\ref{eq:iWW})
the result of the computation of the $W$-box diagram
\cite{Pap94}.

In Eqs.~(\ref{eq:iWW}),
${\cal S}$ is $+1$ if the fermion $f$ has $T_{3f} = -1/2$,
and $-1$ if it has $T_{3f} = 1/2$.
This we may write as ${\cal S} = - 2\ T_{3f}$.
The factor ${\cal S}$ arises from the anti-symmetry
of the three-gauge-boson vertices
under the $W^+ \leftrightarrow W^-$ interchange.

\subsection{The scalar contributions}

We now present the contributions to the MDMs
that involve scalar particles in the loop.
The letters $h$,
$H$ and $A$ refer to the neutral scalars $h^0$,
$H^0$ and $A^0$
of model II,
respectively.
The charged scalars $H^\pm$ are denoted by the letter $C$.
The corresponding Feynman rules may be found,
for example,
in Ref.~\cite{Gun90}.

The formulas we present are for a fermion with $T_{3f} = -1/2$,
such as the bottom or the $\tau$.
For a fermion with $T_{3f} = 1/2$,
such as the top quark,
one must make the following substitutions
\be
\sin{\beta} \leftrightarrow \cos{\beta}
\hspace{2mm} , \hspace{3mm}
\sin{\alpha} \leftrightarrow \cos{\alpha}\, .
\label{eq:substitutions}
\ee

We find
\ba
b^{i C C} &=& {\cal S}\,
\left(\begin{array}{c}
	\cot{(2 \theta_W)}\\
		1
	\end{array}\right)
\frac{1}{s_W^2 M_W^2}
\left\{  \left[
	   (m_f \tan{\beta})^2
		+ (m_i / \tan{\beta})^2
	\right]
	\left(I_{21}-I_{22}-I_{10}\right)
\right.
\nonumber\\
  &  &
\left.
	\hspace{47mm}
	   + m_i^2
	\left( 2 I_{10} - I_{00} \right)
\right\}(m_i,M_C,M_C,q)\, ,
\nonumber\\*[3mm]
b^{C i i} &=&
\frac{1}{s_W^2 M_W^2}
\left\{  \left[
	   (m_f \tan{\beta})^2 (x_i+y_i) +
	(m_i / \tan{\beta})^2 (x_i-y_i)
	\right]
	\left(I_{21}-I_{22}-I_{10}\right)
\right.
\nonumber\\
   &  &
\left.
	\hspace{16mm}
	- 2 x_i m_i^2 I_{10}
\right\}(M_C,m_i,m_i,q)\, ,
\nonumber\\*[3mm]
b^{A f f} &=&
\frac{x_f m_f^2}{s_W^2 M_W^2}\, \tan^2{\beta}\,
\left[I_{21}-I_{22}\right](M_A,m_f,m_f,q)
\label{eq:masterTHDM}\, ,
\nonumber\\*[3mm]
b^{H f f} &=&
\frac{x_f m_f^2}{s_W^2 M_W^2}\,
\frac{\cos^2 \alpha}{\cos^2 \beta}\,
\left[I_{21}-I_{22}-2I_{10}\right](M_H,m_f,m_f,q)\, ,
\nonumber\\*[3mm]
b^{h f f} &=&
\frac{x_f m_f^2}{s_W^2 M_W^2}\,
\frac{\sin^2 \alpha}{\cos^2 \beta}\,
\left[I_{21}-I_{22}-2I_{10}\right](M_h,m_f,m_f,q)\, .
\ea
These diagrams contribute both to the photon and to the $Z$ vertices.

In addition,
there are a few diagrams contributing exclusively to the $Z$ vertex.
In some of them the external $Z$ couples either to $H^0$ and $A^0$,
or to $h^0$ and $A^0$,
in the loop.
Those diagrams do not yield any contribution to the WMDM.
In other diagrams the external $Z$
couples either to a $Z$ and a $H^0$,
or to a $Z$ and a $h^0$,
in the loop (see Fig.~3a).
These diagrams have the same chiral properties
as the MDM operator and,
as a consequence,
their contributions do not get $(m_f/M_W)^2$ pre-factors.
The results are
\ba
b^{f Z H} + b^{f H Z} & = &
\frac{v_f}{s_W^3 c_W^3}\,
(\cos^2{\alpha} + \tan{\beta} \sin{\alpha} \cos{\alpha})
\left[I_{10}+I_{11}\right](m_f,M_Z,M_H,q)\, ,
\nonumber\\*[3mm]
b^{f Z h} + b^{f h Z} & = &
\frac{v_f}{s_W^3 c_W^3}\,
(\sin^2{\alpha} -  \tan{\beta} \sin{\alpha} \cos{\alpha})
\left[I_{10}+I_{11}\right](m_f,M_Z,M_h,q)\, .
\label{eq:masterSpecial}
\ea
To obtain the SM Higgs scalar contributions from the
expressions in this subsection,
one deletes all the diagrams with $H^{\pm}$,
$A^0$,
or $H^0$,
and sets
\be
\beta = \alpha + \pi/2\ ,
\ee
in the diagrams with $h^0$.

\vspace{5mm}

\vspace{5mm}

\newpage

\hspace{10mm} FIGURE CAPTIONS

\vspace{5mm}

Figure 1:
Typical Feynman diagram contributing to the magnetic dipole
moments at one loop.

\vspace{5mm}

Figure 2:
Electroweak one-loop vertex corrections to the MDMs,
excluding the diagrams with the Higgs scalar.

\vspace{5mm}

Figure 3:
One-loop vertex corrections due to the scalars
in the THDM.
The scalars $H^0$,
$h^0$,
$A^0$ and $H^\pm$
are denoted by the letters H,
h,
A and C,
respectively.
For the SM,
only diagrams 3a and 3b exist,
having the Higgs particle as the intermediate neutral scalar.

\vspace{5mm}

Figure 4:
Plot of the contribution of the diagram in
Fig.~3a to the WMDM of the $\tau$,
as a funtion of the mass of the neutral scalar.
This is to be multiplied by
$\cos^2 \alpha + \tan{\beta} \sin{\alpha} \cos{\alpha}$
for $H^0$,
and by
$\sin^2 \alpha - \tan{\beta} \sin{\alpha} \cos{\alpha}$
for $h^0$,
in the THDM.

\vspace{5mm}

Figure 5:
Plot of the contribution of the diagram in
Fig.~3b to the WMDM of the $\tau$,
as a funtion of the mass of the neutral scalar.
This is to be multiplied by
$(1+\tan^2 \beta)$ times
$\cos^2 \alpha$ for $H^0$,
and times $\sin^2 \alpha$ for $h^0$,
in the THDM.

\vspace{5mm}

Figure 6:
Plot of the contribution of the diagram in
Fig.~3c to the WMDM of the $\tau$,
as a funtion of the mass of the neutral pseudo-scalar.
This is to be multiplied by
$\tan^2 \beta$.

\vspace{5mm}

Figure 7:
Plot of the sum of the contributions of the diagrams in
Figs.~3d and 3e to the WMDM of the $\tau$,
as a funtion of the mass of the charged scalar.
This is to be multiplied by
$\tan^2 \beta$.

\vspace{5mm}

Figure 8:
Gluonic one-loop vertex correction to the MDMs of the quarks.

\end{document}